# Lower bounds for the maximum number of runners that cause loneliness, and its application to Isolation

Deepak Ponvel Chermakani
deepakc@pmail.ntu.edu.sg   deepakc@e.ntu.edu.sg   deepakc@ed-alumni.net   deepakc@hawaii.edu   deepakc@myfastmail.com
deepakc@usa.com

**Abstract: -** We consider (n+1) runners with given constant unique integer speeds running along the circumference of a circle whose circumferential length is one, and all runners starting from the same point. We define and give lower bounds to a first problem PMAX of finding, for every runner r, the maximum number of runners that can be simultaneously separated from runner r by a distance of atleast d. For $d=1/(2^{\lfloor \lg(n) \rfloor})$, a lower bound for PMAX is $(\,n - ((n-1)/\lfloor \lg(n) \rfloor)\,)$, which makes the fraction of simultaneously separated runners tend to 1 as n tends to infinity. Next, we define and give upper bounds to a second problem ISOLATE of finding, for every runner r, the minimum number of steps needed to isolate r, assuming that the runners that can be simultaneously separated from r by atleast d, are removed at each step. For $d=1/(2^{\lfloor \lg(n) \rfloor})$, an upper bound for ISOLATE is $(\,\lg(n-1) / \lg(\lfloor \lg(n) \rfloor)\,)$.

## 1. Introduction

Given (n+1) points (called runners) with unique constant integer speeds moving along the circumference of a circle of unit length of circumference, starting from the same point at time t=0, a runner is said to be lonely at a time t if it is simultaneously separated from all other n runners by a distance of atleast 1/(n+1) at t. The well known Lonely Runner Conjecture (LRC) is that for every runner r, there exists a time $t_r$ at which r becomes lonely [1][2][3].

Though inspired from the LRC, this paper defines and focuses on two different problems.

The first problem PMAX is to find, for every runner r, the lower bound for the maximum number of runners that can be simultaneously separated from r by a distance $\geq$ d, where d is a given constant. Clearly, the answer to PMAX =n for d=(1/(n+1)), if and only if, the LRC is true.

The second problem ISOLATE is to find, for given distance d, for every runner r, the upper bound for the minimum number of steps needed to isolate runner r, assuming that the runners that can be simultaneously separated from runner r by a distance $\geq$ d, are removed at each step.

## 2. Approach

### 2.1 Notations and Definitions

We denote:
1. the total number of runners, including runner r, as (n+1)
2. runner r as simply runner 0.
3. G as the set of runners other than runner 0.
4. each runner initially in G, as a unique integer in [1, n].
5. the speed of runner 0 as 0 (i.e. $s_0 = 0$), and the integer speeds of the other runners (i.e. $s_i$ for each runner i in [1,n]) as the absolute value of their corresponding relative speeds with respect to $s_0$. This approach is inspired from papers on the LRC [1][2][3].
6. runner 0's position as permanently fixed at zero, conveniently denoted as the top of the circle.
7. the distance between any two runners i and j, as the shortest distance along the circumference of the circle between runners i and j. For example, if at time t, runner i is located at the left most point of the circle, and since runner 0 is fixed at the top of the circle, we say that the distance between runner i and runner 0 is 0.25 at time t, and not 0.75.
8. d as a given constant that is a negative integer power of two, where $0 < d \leq 0.5$.
9. for each runner i in [1, n], $E_i$ as the positive integer representing the position (also called index) of the least significant "1" of $s_i$ from the right most 0 bit of $s_i$, when $s_i$ is written in binary format. For example, if n=3, $s_1$=2,

$s_2=7$, $s_3=16$ and $s_0=0$, then in binary format, $s_1=0010$, $s_2=0111$, $s_3=1000$, so $E_1 = 2$, $E_2 = 1$ and $E_3 = 4$.
10. p as max($E_i$, over all runners i in [1, n]). For the previous example, p = 4.
11. any time t as a sum of some non-negative integer plus a fraction that is the sum of negative integer powers of two, i.e. $t = NNI + b_1\ 2^{-1} + b_2\ 2^{-2} + b_3\ 2^{-3} + ... + b_{p-1}\ 2^{-p+1} + b_p\ 2^{-p}$, where $b_j \in \{0, 1\}$ for each integer j in [1, p], and where NNI denotes some non-negative integer. This is the notation we shall follow when we try to prove the existence of time t in this paper. Note that when t is a positive integer, all n runners are at the same position on the circle as the position they are at t=0, as they all have integer speeds. The existence of a vector $<b_p, b_{p-1}, b_{p-2}, ..., b_2, b_1>$ that satisfies the conditions of any Theorem in our paper obviously implies the existence of some real time t satisfying that Theorem, though we know that the converse is not necessarily true.
12. using the above definition of t, the position of any runner i at time t, as the fractional part of ($t\ s_i$).
13. the status of a runner i as SAVED if it is separated from runner 0 by a distance $\geq d$, given the current values of the binary variable vector $<b_p, b_{p-1}, b_{p-2}, ..., b_2, b_1>$.
14. the status of a runner i as UNSAVED, if it is not SAVED, given the current values of the binary variable vector $<b_p, b_{p-1}, b_{p-2}, ..., b_2, b_1>$.
15. floor(x) as the greatest integer smaller than x.
16. ceiling(x) as the smallest integer greater than x.
17. lg(x) as the logarithm of x to the base 2.
18. a^b as $a^b$, and we shall use both notations where convenient.
19. LCM as Least Common Multiple.

## 2.2 Problem PMAX

We first state and prove a trivial Theorem 1.

**Theorem 1:** For each runner i of G, there exists a time $t_i$ at which runner i is separated from runner 0 by d = 0.5.
**Proof:** The $t_i$ defined by setting the $b_j = 1$ for $j = E_i$, and setting the $b_j = 0$ for all integers j in [1, p] and $j \neq E_i$, puts the position of runner i at 0.5, while runner 0 continues to remain at 0. At this $t_i$, the runners k, where $k \neq i$ and $k \neq 0$, can be anywhere on the circumference of the circle.
**Hence Proved Theorem 1.**

Next, we state and prove Theorems 2 and 3, which respectively give some lower bounds for PMAX, for d = 0.25 and d = $1/(2^{floor(lg(n))})$.

**Theorem 2:** There exists a time t at which atleast ceiling(n/2) runners of G are simultaneously separated from runner 0 by $d \geq 0.25$.
**Proof:** The time t can be defined by the following algorithm:
1. Initialize $b_j = 0$ for all integers j in [1, p], ensuring that the position of all runners is at 0 with runner 0.
2. Sort the runners i of G in descending order of $E_i$, and denote this sorted list as L. If two or more runners have the same $E_i$, they occupy the same position in the list. For example, L could look like this for n=6:
   Item 1 of L: runner 3, runner 6
   Item 2 of L: runner 1, runner 2, runner 4
   Item 3 of L: runner 5
3. $k = E_i$ of the runners of the topmost item of L.
4. For $j = E_k$, choose $b_j = 1$ or 0, to ensure that atleast half of the runners of the topmost item of L would be SAVED. Since the SAVED runners are ensured to be diametrically opposite on the circumference from their UNSAVED position, the SAVED runners are SAVED by a distance of atleast 0.25.
5. For every runner i in [1, n], update (position of runner i) = fraction((position of runner i) + ($s_i\ b_k\ 2^{-k}$)).
6. Remove the topmost item from L.
7. Go to step 3, if L is not empty.

Note that since L was sorted in descending order of $E_i$, step 4 (i.e.setting the $b_j$) for any of the later items of L will not change the position (and hence will not affect the SAVED status or UNSAVED status) of any of the runners in earlier items of L.
**Hence Proved Theorem 2.**

**Theorem 3:** There exists a time t at which atleast ( n - ((n-1)/floor(lg(n))) ) runners of G are simultaneously separated

**from runner 0 by d $\geq 1/(2^{\text{floor}(\lg(n))})$.**
**Proof:** We prove Theorem 3 using Lemmas 3.1, 3.2 and 3.3:

<u>Lemma 3.1</u>: Denote the index of the left most column of a movable window W of (c-1) columns as integer k, such that (c-1) $\leq k \leq$ (p+c-2). Then the following three statements are true:
1. Denote the set of runners whose $E_i$s are such that $k \geq E_i \geq$ (k-c+2), as $S_k$. There exists a value of $b_k$ (= 1 or 0) that will SAVE atleast half of the runners of $S_k$ by a distance $\geq 1/2^c$.
2. After $b_k$ is set to save atleast half of the runners of $S_k$ by a distance $\geq 1/2^c$, by positioning them in odd numbered sectors, the status of these SAVED runners i of $S_k$ will not get changed from SAVED to UNSAVED by the choice of any $b_j$ as 1 or 0, if $k > j \geq E_i$.
3. The choice of any $b_j$ as 1 or 0, will change the status of any runner i of $S_k$, neither from SAVED to UNSAVED, nor from UNSAVED to SAVED, if $E_i > j$.

<u>Proof</u>: Consider any runner i of $S_k$. Divide the circle into $2^u$ sectors where u = (k - $E_i$ + 1), numbered from q=0 to q=($2^u$ - 1), with the arc of sector q being from (-1/$2^{u+1}$ + q/$2^u$) to (+1/$2^{u+1}$ + q/$2^u$). If one value of $b_k$ sets the position of runner i in an odd numbered sector, then the other value of $b_k$ sets the position of runner i in an even numbered sector. Similarly, if one value of $b_k$ sets the position of atleast half of the runners of $S_k$ in odd numbered sectors, then the other value of $b_k$ sets the position of those same runners (i.e. atleast half of the runners of $S_k$) in even numbered sectors. We set the value of $b_k$ to ensure that the position of atleast half of the runners of $S_k$ are in odd numbered sectors, which will automatically ensure that these runners are SAVED by a distance $\geq 1/2^{u+1} \geq 1/2^c$, since the maximum value of u = (c-1). This proves the first statement.

Consider any runner i of $S_k$ that has been SAVED to arrive at an odd numbered sector, after setting $b_k$ to SAVE atleast half of the runners of $S_k$. The choice of any $b_j$ as 1 or 0, for $k > j \geq E_i$, will only have the effect of moving runner i from one odd numbered sector to another odd numbered sector, which will again ensure that runner i is SAVED by a distance $\geq 1/2^{u+1} \geq 1/2^c$. This proves the second statement.

The third statement of this Lemma is trivial because for all j < $E_i$, any value of $b_j$ will contribute to a 0 change in the position of runner i. Hence, there will be no change in the SAVED status or UNSAVED status of runner i.
<u>Hence Proved Lemma 3.1</u>.

<u>Lemma 3.2</u>: For each integer c in [2, floor(lg(n))], there exists a time $t_c$ at which no more than ((n - 1)/c) runners of G are simultaneously separated from runner 0 by a distance < $1/2^c$.
<u>Proof</u>: Consider a window W of (c-1) columns, initially placed such that W's right most column is aligned with column p. For the same previous example of n=3, $s_1$=2, $s_2$=7, $s_3$=16 and $s_0$=0, the initial position of W is such that the rightmost column of W coincides with column number 4.
Now perform the steps of the following algorithm:
1. Initialize $b_j$ = 0 for all integers j in [1, p+c-2], and mark the status of all runners of G as UNSAVED.
2. Initially place W such that the index of W's right most column is aligned with column p. That is k = (p+c-2), where k denotes the index of W's left most column.
3. Create set S = set of runners i, whose $E_i$ is such that k $\geq E_i \geq$ (k-c+2), and whose status is not marked as SAVED.
4. Choose $b_k$ = 0 or 1, to position ceiling(sizeof(S)/2) runners of S, in odd numbered sectors as defined in Lemma 3.1, and mark their status to SAVED.
5. For every runner i of G, update (position of runner i) = fraction((position of runner i) + ($s_i\ b_k\ 2^{-k}$)).
6. Move W one column to the right, meaning that k = k - 1.
7. Go to step 3, if k $\geq$ min($E_i$, over all runners i of G, whose status is not marked as SAVED).

We shall now prove that, even in the worst case, the number of UNSAVED runners after the above algorithm terminates, is $\leq$ (n - 1)/c. There are two conditions that need to be satisfied at each iteration of the above algorithm (i.e. each movement of W), to maximize the number of UNSAVED runners until the algorithm terminates:
1. This condition pertains to the worst case performance of the algorithm. In this condition, during each movement of W, Step 4 SAVES ceiling(sizeof(S)/2) runners of S with the least $E_i$, placing them in odd numbered sectors. This maximizes the number of potential UNSAVED runners whose ending "1" bit is at the left of the moving window W, so that they can escape from W sooner. Note that if Step 4 SAVES any other set of runners, the ending "1" bits of the speeds of the potential UNSAVED runners would have to pass through a greater (or equal) number of columns of W, resulting in a greater (or equal) number of them being SAVED.
2. This condition pertains to the worst case configuration of the ending "1" bits of the speeds of the runners, along with the worst case performance of the algorithm. In this condition, during each movement of W, the $E_i$ of the potential

UNSAVED runners remain in a single column (this will appear as an UNSAVED column moving leftwards through W, as W moves rightwards one column at a time). Note that if the UNSAVED runners are scattered among multiple columns of W, there would be a greater (or equal) number of columns of W that the ending "1" bits of the speeds of the UNSAVED runners would have to pass through, resulting in a greater (or equal) number of them being SAVED.

One choice for maximizing the number of UNSAVED runners after the execution of the algorithm is as follows:
1. The unique values of the $E_i$ of the runners have to be in sequence with no gap, when the $E_i$ are arranged in ascending or descending order. For example, we could have the $E_i$ of runners as 7, 6, 6, 6, 5, 5, 4, 4, 4, 3, 2, 2. The reason is that the existence of any gap will cause more runners to be SAVED, especially for those runners whose $E_i$ is after the gap.
2. Exactly 1 runner i has to have $E_i = p$ (i.e. the largest $E_i$), since all the runners with the largest $E_i$ can be SAVED, simply by setting $b_j = 1$ for $j=E_p$.
3. Column (p-1) has to be the column where all the UNSAVED runners (at the end of the algorithm) have to be concentrated. Since the number of runners is halved once column (p-1) crosses the rightmost column of W, we denote the initial number of runners i with $E_i = (p-1)$ as 2x, where x would be the final number of UNSAVED runners.
4. The number of runners with $E_i$ in each column right of column (p-1) has to be = to half of the number of runners in column (p-1) = 0.5(2x) = x. This will ensure that the number of UNSAVED runners in column (p-1) continues to remain x until W crosses column (p-1) completely.

Following the above mentioned strategy, one choice for the maximum number of runners x that can escape without being separated by $d \geq 1/2^c$ is given by the following configuration:
Number of {runners i such that $E_i = p$} = 1
Number of {runners i such that $E_i = p-1$} = 2x
Number of {runners i such that $E_i = p-2$} = x
Number of {runners i such that $E_i = p-3$} = x
Number of {runners i such that $E_i = p-4$} = x
...
Number of {runners i such that $E_i = p-c+1$} = x
which implies that:
$1 + 2x + x(c - 2) = n$, which implies that:
$x = (n - 1)/c$, which completes the proof, under the assumption that $(n - 1)/c$ is an integer. If $(n - 1)/c$ is not an integer, then the upper bound on x is given by $x \leq floor((n - 1)/c)$, since a greater number of runners would be saved.

Note that there are other possible global maxima for x. The same global maximum on x is obtained by having a stream of 1, 2, 1, 1, .....(c-2) times, 2, 1, 1, .....(c-2) times, ... , which gives
$1 + x(2 + c - 2) = n$, which again gives
$x \leq floor((n - 1)/c)$.

The same global maximum on x is obtained by having a stream of 1, 4, 2, 2, .....(c-2) times, 4, 2, 2, .....(c-2) times, ...
$1 + (x/2)(4 + 2(c - 2)) = n$, which again gives
$x \leq floor((n - 1)/c)$.

In a similar way, there are multiple other global maxima.
Hence Proved Lemma 3.2.

Lemma 3.3: For each integer c in [2, floor(lg(n))], there exists a time $t_c$ at which atleast ( n - ((n - 1)/c) ) runners of G are simultaneously separated from runner 0 by $d \geq 1/2^c$.
Proof: From Lemma 3.2, there exists a time $t_c$ at which not more than floor((n - 1)/c) runners are simultaneously separated from runner 0 by a distance $\leq 1/2^c$. It follows trivially that at this same time $t_c$, the remaining atleast ( n - floor((n - 1)/c) ) runners of G are simultaneously separated from runner 0 by a distance $\geq 1/2^c$.
Hence Proved Lemma 3.3.

Substituting c=floor(lg(n)) in Lemma 3.3, Theorem 3 follows.
**Hence Proved Theorem 3.**

**Theorem 4:** As n→∞, there exists a time t at which the fraction of runners of G that can be simultaneously separated

**from runner 0 by d ≥ 1/n, tends to 1.**
**Proof:** From Theorem 3, since $n \geq 2^{floor(lg(n))}$ for all positive integers n, there exists a time t at which the fraction of runners of G that can be simultaneously separated from runner 0 by a distance $\geq 1/2^{floor(lg(n))} \geq 1/n$, is $\geq ( n - (n - 1)/floor(lg(n)) ) / n$. As n→∞, this fraction tends to (1 - 1/floor(lg(n)) ), which tends to 1.
**Hence Proved Theorem 4.**

## 2.2   Isolation of runner 0

We now proceed to the problem ISOLATE, which we defined earlier. We give upper bounds to four versions of ISOLATE, respectively for four values of d.
   Theorems 5, 6 and 7, respectively, give some upper bounds to ISOLATE for d = 0.5, d = 0.25 and $d = 1/(2^{floor(lg(n))})$. Theorem 8 gives an upper bound on the minimum number of steps to ISOLATE for $d = 1/n_m$, where $n_m$ is the total number of runners (excluding runner 0) remaining at the $m^{th}$ step.

**Theorem 5:** Runner 0 can be isolated in a number of steps ≤ n, where each step consists of removing the SAVED runners from G that can be simultaneously separated from runner 0 by d = 0.5.
**Proof:** Consider the following algorithm:
   1. Initialize $b_j = 0$ for all integers j in [1, p], ensuring that the position of all runners is at 0 with runner 0.
   2. Sort the runners i of G in ascending order of $E_i$, and denote this sorted list as L. If two or more runners have the same $E_i$, they occupy the same position in the list. For example, L could look like this for n=6:
      Item 1 of L: runner 3, runner 6
      Item 2 of L: runner 1, runner 2, runner 4
      Item 3 of L: runner 5
   3. k = $E_i$ of the runners of the topmost item of L.
   4. For j=$E_k$, set $b_j = 1$.
   5. Update (position of runners i of topmost item of L) = 0.5.
   6. Remove runners i of topmost item of L, from both L and G.
   7. Goto step 3 if G is not empty.
Clearly, each pass through steps 5 and 6 removes atleast 1 runner simultaneously separated from runner 0 by a distance of 0.5. So the maximum number of passes is n.
**Hence Proved Theorem 5.**

**Theorem 6:** Runner 0 can be isolated in a number of steps ≤ lg(n), where each step consists of removing the SAVED runners from G that can be simultaneously separated from runner 0 by d ≥ 0.25.
**Proof:** Consider the following algorithm:
   1. Initialize $b_j = 0$ for all integers j in [1, p], ensuring that the position of all runners is at 0 with runner 0.
   2. From Theorem 2, there exists a vector $<b_p, b_{p-1}, b_{p-2}, ... , b_2, b_1>$ at which atleast half of the remaining runners of G would be simultaneously separated from runner 0 by a distance ≥ 0.25 (denote these runners as SAVED runners).
   3. Using the above $\underline{b}$ vector, for every runner i of G, update (position of runner i) = fraction( (position of runner i) + ($s_i$ SUMMATION(($b_j 2^{-j}$), over all integers j in [1,p])) ).
   4. Remove the SAVED runners from G.
   5. Goto step 1 if G is not empty.
If we denote $n_m$ as the number of UNSAVED runners remaining after step m, then from Theorem 2, $n_{m+1} \leq floor(n_m / 2)$, where $n_1 = floor(n/2)$. Thus the minimum number of steps to isolate runner 0 is upper bounded by the number of steps needed for the sequence $n_{m+1} = floor(n_m / 2)$ to reach 0, where $n_1 = floor(n/2)$. Clearly, this upper bound = lg(n).
   Note that the use of this recurrence is equivalent to the loop from Step 5 back to Step 1 where the b vector  (i.e. time vector) is reset to 0, and is valid since the positions of all runners will periodically merge with runner 0 at the initial starting point with period = $LCM(1/s_1, 1/s_2, 1/s_3, ... , 1/s_n)$, one choice of which is = $PRODUCT(s_1, s_2, s_3, ... , s_n) = PRODUCT(s_i$, over all integers i in [1, n] ).
**Hence Proved Theorem 6.**

**Theorem 7:** Runner 0 can be isolated in a number of steps ≤ lg(n - 1)/lg(floor(lg(n))), where each step consists of removing the SAVED runners from G that can be simultaneously separated from runner 0 by $d \geq 1/(2^{floor(lg(n))})$.
**Proof:** If we denote $n_m$ as the number of UNSAVED runners remaining after step m, then from Theorem 3, $n_{m+1} \leq floor((n_m$

- 1)/floor(lg(n))). Since our aim is to find an upper bound U on the number of steps, we aim to find the value of m at which the recurrence sequence $n_{m+1}$ = floor(($n_m$ - 1)/floor(lg(n))), becomes equal to 1, which is lesser than the number of steps taken for the sequence $n_{m+1}$ = ($n_m$ - 1)/floor(lg(n)) to become = 1. So the upper bound U is given by (n - 1)/(floor(lg(n)))$^U$ = 1, which gives U = lg(n - 1)/lg(floor(lg(n))).
**Hence Proved Theorem 7.**

**Theorem 8:** Where each step consists of removing the SAVED runners from G that can be simultaneously separated from runner 0 by d $\geq$ 1/(2^floor( lg($n_m$) )), where $n_m$ is the number of runners excluding runner 0 remaining in the m$^{th}$ step, then runner 0 can be isolated in a number of steps < ( 2 + ((N - 2$^L$)/L) + SUMMATION((2$^k$/k), over all integers k in [1, L-1]) ), where N = (lg(n) - 1), and L = floor(lg(N)) = floor(lg(lg(n) - 1)).
**Proof:** From Theorem 3, we have $n_{m+1} \leq$ floor(($n_m$ - 1)/floor(lg($n_m$))). So, we can write for an upper bound:
$n_{m+1} \leq$ ($n_m$ - 1)/(lg($n_m$) - 1), which is equivalent to
$n_{m+1} \leq$ ($n_m$ - 1)/lg($n_m$/2), which is equivalent to
($n_{m+1}$/2) $\leq$ (($n_m$/2 ) - 0.5)/lg($n_m$/2)
Denoting $M_m$ = ($n_m$/2) , we get
$M_{m+1} \leq$ ($M_m$ - 0.5) / lg($M_m$)

Since our aim is to find an upper bound on the minimum number of steps in which runner 0 can be isolated, we aim to find the number of steps for the recurrence sequence $n_{m+1}$ = ($n_m$ - 1)/(lg($n_m$) - 1) to become $\leq$ 4 (after which an extra 2 steps would be needed to isolate runner 0), which is equal to the number of steps for the sequence $M_{m+1}$ = ($M_m$ - 0.5) / lg($M_m$) , to become $\leq$ 2, which is lesser than the number of steps for the sequence $M_{m+1}$ = $M_m$ / lg($M_m$), to become $\leq$ 2. Taking logs on both sides, we have the recurrence lg($M_{m+1}$ ) = lg($M_m$) - lg(lg($M_m$)). Denoting $N_m$ = lg($M_m$), we have $N_{m+1}$ = $N_m$ - lg($N_m$ ). And our aim is now to find an upper bound on the number of steps for this sequence to become $\leq$ 1. One upper bound can be found by finding an upper bound for the number of steps $S_N$ taken for the sequence $N_{m+1}$ = $N_m$ - floor(lg($N_m$ )) to become $\leq$ 1. This upper bound $S_N$ is given by: $S_N$ = ((N - 2$^L$)/L) + (2$^{L-1}$/(L-1)) + (2$^{L-2}$/(L-2)) + ... + (2$^3$/3) + (2$^2$/2) + (2$^1$/1), where N = (lg(n) - 1), and L = floor(lg(N)) = floor(lg(lg(n) - 1)). Add 2 (reason mentioned earlier in proof) to get the Theorem.
**Hence Proved Theorem 8.**

## 3.  Future Work

Variants and special cases of PMAX and ISOLATE could be studied in future. Future work could be to extend similar bounds to PMAX and ISOLATE, for d being a given general rational > 0 and < 0.5, as our paper has currently only considered d to be a given negative integer power of 2. As a first step, our Theorems could be extended to cover cases of d being a given sum of negative integer powers of 2 that could approximate a rational.

Another future work could be to apply specific Theorems in our paper to prove special cases of the LRC. With regard to attempts to prove the LRC in general (assuming the LRC is true), our paper seems to suggest that the definition of time t as a sum of negative integer powers of 2, might not help, as some runners would always escape UNSAVED. There is also difficulty in expressing certain values of d like 1/(n+1) as the sum of negative integer powers of two if (n+1) is prime. Hence, for the LRC, t might need to be defined as a real (or atleast a rational) variable.

**References**
[1] T. Tao, "Some remarks on the lonely runner conjecture", arXiv:1701.02048v4, Nov 2017.
[2] J. Barajas, O. Serra, "The lonely runner with seven runners", Electronic Journal of Combinatorics, 15, R48, 18 pp, 2008.
[3] W. Bienia, "Flows, view obstructions and the lonely runner", Journal of Combinatorial Theory, Article number TB971770, 1998.

**About the author**
I, Deepak Ponvel Chermakani, wrote this paper out of my own interest and initiative, during my spare time. In Aug 2015, I completed a fulltime two-year Master of Science Degree in Electrical Engineering, from University of Hawaii at Manoa USA (www.hawaii.edu). In Sep 2010, I completed a fulltime one-year Master of Science Degree in Operations Research with Computational Optimization, from University of Edinburgh UK (www.ed.ac.uk). In Jul 2003, I completed a fulltime four-year Bachelor of Engineering Degree in Electrical and Electronic Engineering, from Nanyang Technological University Singapore (www.ntu.edu.sg). In Jul 1999, I completed fulltime high schooling from National Public School in Bangalore in India. I am most grateful to my parents (especially my mother Mrs. Kanaga Rathinam Chermakani) for their sacrifices in educating me and bringing me up.